# Predicting Endocrine Disruptors: A Deep Learning QSAR Model for Estrogen Receptor Activity

Belaguppa Manjunath Ashwin Desai
School of Engineering
Dayananda Sagar University
Bengaluru, India
ashwin.desai@ieee.org

Shreyas Murthy
School of Engineering
Dayananda Sagar University
Bengaluru, India
shreyasmurthy856@gmail.com

Bhoomika Sridhar
School of Basic and Applied Sciences
Dayananda Sagar University
Bengaluru, India
bhoomi.sree03@gmail.com

Anirudh Belaguppa Manjunath
Khoury College of Computer Sciences
Northeastern University
Boston, USA
anirudhdesai777@gmail.com

Vivien Humtsoe
School of Basic and Applied Sciences
Dayananda Sagar University
Bengaluru, India
humtsoevivien@gmail.com

Pronama Biswas
School of Basic and Applied Sciences
Dayananda Sagar University
Bengaluru, India
pronama-sbas@dsu.edu.in

***Abstract*— Endocrine-disrupting chemicals (EDCs) threaten human health, ecosystems, and biodiversity by interfering with hormonal signaling pathways conserved across vertebrates. Traditional in vivo assays are costly and time-consuming, limiting their capacity to screen the growing number of chemicals. To address this, we developed a deep learning-based QSAR model to predict estrogen receptor (ER) binding molecules. Using a curated dataset of 224 compounds and 2,944 molecular descriptors and fingerprints, a deep neural network (DNN) incorporating dropout and batch normalization was trained and validated. The model achieved training and test accuracies of 96.65% and 91.30%, respectively, with an ROC-AUC of 0.81, a precision of 0.82, and a recall of 0.88 for the active class. Molecular docking against estrogen receptor (PDB ID: 5TOA) confirmed that several predicted compounds exhibited binding comparable to Estradiol, sharing key interactions. This model enables rapid screening of potential EDCs, supporting efficient chemical risk assessment and contributing to biodiversity conservation by identifying compounds that may disrupt reproduction and population stability in humans and wildlife.**

## I. Introduction

Endocrine-disrupting chemicals (EDCs) constitute a significant and growing global health challenge, impacting both human and environmental health through interference with the hormone system [1]. EDCs, which are ubiquitous in pesticides, plastics, personal care products, and industrial waste, are associated with a wide spectrum of adverse effects, including reproductive disorders, developmental deficits, and hormone-sensitive cancers [1], [2]. Beyond human health, EDCs also threaten biodiversity by altering reproductive success, survival, and population dynamics across a wide range of species, thereby contributing to ecosystem imbalance and biodiversity loss.

The core issue lies in the ability of EDCs to disrupt the endocrine system's homeostasis, the crucial process of maintaining a stable internal environment through tightly regulated hormonal feedback loops. EDCs interfere with this regulation by acting as agonists (mimicking natural hormones like estrogen) or antagonists (blocking the natural hormone from binding to its receptor), hence inappropriately altering signaling pathways [3]. Natural estrogen is the primary hormone controlling the female reproductive system and bone density, its signaling pathway is particularly susceptible to exogenous interference [4].

The current gold standard for assessing the toxicity and endocrine-disrupting potential of chemicals is through in vivo testing, encompassing comprehensive assays such as the Uterotrophic Assay, Hershberger Assay, and the Two-Generation Reproductive Toxicity Study. This methodology is, however, both time-consuming and prohibitively expensive [5]. Given the rapid pace of industrial chemical production and the vast number of compounds already lacking comprehensive toxicological data, traditional testing methods are insufficient for modern risk assessment.

To address these critical limitations, computational toxicology offers a powerful and cost-effective approach for high-throughput chemical screening. Quantitative Structure-Activity Relationship (QSAR) models, which establish mathematical relationships between a compound's molecular structure and its biological activity, are ideally suited for this purpose [6].

A key mechanism of EDC activity is the binding and modulation of nuclear hormone receptors, particularly the estrogen receptor (ER). The Endocrine Disruptor Knowledge Base (EDKB) Estrogen Receptor Binding Dataset was specifically created as a diverse and balanced resource for developing reliable predictive models. This dataset includes 131 established ER binders and 101 non-binders, providing an essential foundation for modeling [7]. To maximize predictive accuracy for this complex endpoint, a deep learning methodology utilizing a deep forward neural network(DNN) was employed. This model processes a large vector of calculated molecular descriptors and fingerprints obtained using RDKit.

The objective of this study is to develop and validate a deep learning–based QSAR model that can accurately classify compounds as estrogen receptor (ER) binders or non-binders. This computational approach (in-silico screening) provides a sustainable and scalable alternative to animal-based testing, enabling early identification of chemicals with potential endocrine-disrupting effects. By reducing the likelihood of releasing harmful ER-active compounds into the environment, the model directly supports biodiversity conservation and aligns with the United Nations Sustainable Development Goals (SDGs), particularly SDG 6 (Clean Water and Sanitation), SDG 14 (Life Below Water), and SDG 15 (Life on Land).

## II. Methods

### A. Data Collection and Pre-processing

The predictive model utilized the established Estrogen Receptor (ER) Binding Database from the U.S. FDA's National Center for Toxicological Research (NCTR) [7]. This resource is recognized as the largest published ER binding database of same-assay results generated in a single laboratory, featuring compounds tested in a well-validated in vitro rat uterine cytosol ER competitive-binding assay. The initial dataset was named “ER_DATASET.csv”. The initial file consisted of 232 chemical compounds selected a priori for structural and activity diversity. Three essential parameters were extracted: SMILES, Activity label, and Compound ID (CID). The dataset underwent a cleaning process to ensure data quality. First, entries with missing values in any of the three columns were filtered out. To maintain the high confidence and fidelity of the model, the compounds classified as "Inconclusive" were systematically removed.

After these refinements, the final curated dataset, was saved as “ER_DATASET_CLEAN.csv”. For binary classification, the activity labels were converted to numerical values, active compounds were assigned a value of 1, and inactive compounds were assigned a value of 0.

### B. Descriptors and Fingerprint Calculation

To convert the canonical SMILES structures into quantitative input for the deep learning model, comprehensive molecular features were generated using RDKit (https://www.rdkit.org/). 217 physicochemical and structural molecular descriptors were computed, capturing essential properties of the compounds. Invalid molecules were identified and subsequently removed to maintain data accuracy. In addition, 5 molecular fingerprints including Morgan, Avalon, MACCS keys, topological torsion, and atom pair fingerprints were generated to assess structural characteristics. These fingerprints and descriptors were combined into a single dataset and saved as “EDC_RDkit_Descriptors.csv” for further modeling. In total, the generated descriptors and combined fingerprints provided 2,944 features for the QSAR deep learning model, ensuring maximal structural information was available for accurate classification.

### C. Model Training and Testing

The compound matrix classified by its activity, was prepared for modeling by first addressing class imbalance. The minority class was oversampled using the resample method from the Scikit-learn library to ensure a balanced representation of active and inactive compounds. The resulting balanced dataset was then partitioned into training (80%) and external test (20%) sets using stratified sampling to maintain consistent class distribution across both partitions. For the classification task, a deep feedforward neural network (DNN) was implemented using the TensorFlow framework. The architecture comprised four fully connected layers: the input layer received the molecular features, followed by two hidden layers containing 64 and 32 neurons, respectively. Each hidden layer incorporated the ReLU activation function, Batch Normalization, and Dropout regularization. The final output layer used a single neuron with a Sigmoid activation function for binary classification. “He_normal” initialization was employed for weight initialization to promote stable and improved convergence during training.

The model was trained for 150 epochs using a batch size of 32. The optimization objective was the minimization of the Binary Cross-Entropy (BCE) loss function, achieved through the Adam optimizer with a fixed learning rate of 0.0004 and weight decay of 0.003. The structured training approach involved passing the batch data through the network, computing the loss against the actual activity labels, and updating the model weights via backpropagation and gradient descent. To enhance generalization and mitigate overfitting, dropout regularization and batch normalization were integral components of the network architecture.

### D. Tools and Libraries

This study was executed using the Python programming environment, leveraging specialized libraries for all computational tasks. RDKit (https://www.rdkit.org/)) was utilized for molecular descriptor and fingerprint generation, directly quantifying the chemical structures. Scikit-learn (https://scikit-learn.org)) managed crucial steps in data pre-processing, including feature scaling and dataset splitting, in addition to supporting performance evaluation. The deep learning experiments and neural network training were implemented using TensorFlow (https://www.tensorflow.org). Efficient numerical operations and data handling were executed using the NumPy and Pandas libraries. Finally, results were visualized using Matplotlib (https://matplotlib.org) and Seaborn (https://seaborn.pydata.org) to generate graphical representations, such as ROC curves and confusion matrices. Computational tasks were performed on a high-performance MacBook Air equipped with an M2 core processor and 16 GB of RAM, running Mac OS Sequoia version 15.5. To ensure dependency management and reproducibility, all steps from dataset preparation through model training and evaluation were executed within a dedicated Conda environment.

### E. Dataset for Prediction

For external validation and demonstration of real-world applicability, a dedicated dataset of 100 compounds was curated for prediction, named “pred_edc.csv”. This collection was assembled from the PubChem database and specifically selected to represent chemicals to which living organisms are regularly exposed in daily life, comprising a balanced mixture of known EDCs and non-EDC compounds. The canonical SMILES notation and PubChem Compound ID (CID) were extracted, verified, and utilized as the structural basis for feature calculation. The previously trained DNN model was applied to this external dataset to obtain predictive probabilities for ER binding. A classification threshold of 0.5 was implemented: compounds with a predicted probability greater than or equal to 0.5 were classified as Active (1), while those below 0.5 were classified as Inactive (0).

### F. Molecular Docking Methods

To validate the predicted compounds, molecular docking experiments were performed [6], [8], with binding affinities compared against Estradiol (CID: 5757) [9]. The Estrogen Receptor protein structure, chain B (PDB ID: 5TOA) was retrieved from the RCSB Protein Data Bank and its quality was evaluated using PROCHECK Ramachandran Plot tools (https://saves.mbi.ucla.edu/), ensuring that selected models achieved a score above 90%. Prior to docking, the stereochemistry of each SMILES structure was corrected and converted into SDF format, followed by conversion into PDBQT files with Gasteiger charges applied using Open Babel. Preparation of the Estrogen Receptor protein and Estradiol reference ligand was carried out in AutoDockTools 1.5.7 (https://autodock.scripps.edu/) and saved in PDBQT format [6], [10]. Docking simulations were executed in a Conda environment with a Vina script, accessed through Visual Studio Code. Predicted compounds were subjected to the same docking procedure [9]. The resulting protein–ligand complexes were analyzed for binding sites and interacting residues using UCSF Chimera 1.17 (https://www.cgl.ucsf.edu/chimera/) and Discovery Studio 2021 Client (https://www.3ds.com/products/biovia/discovery-studio/visualization) [11]. A schematic overview of the methodology is presented in Fig. 1.

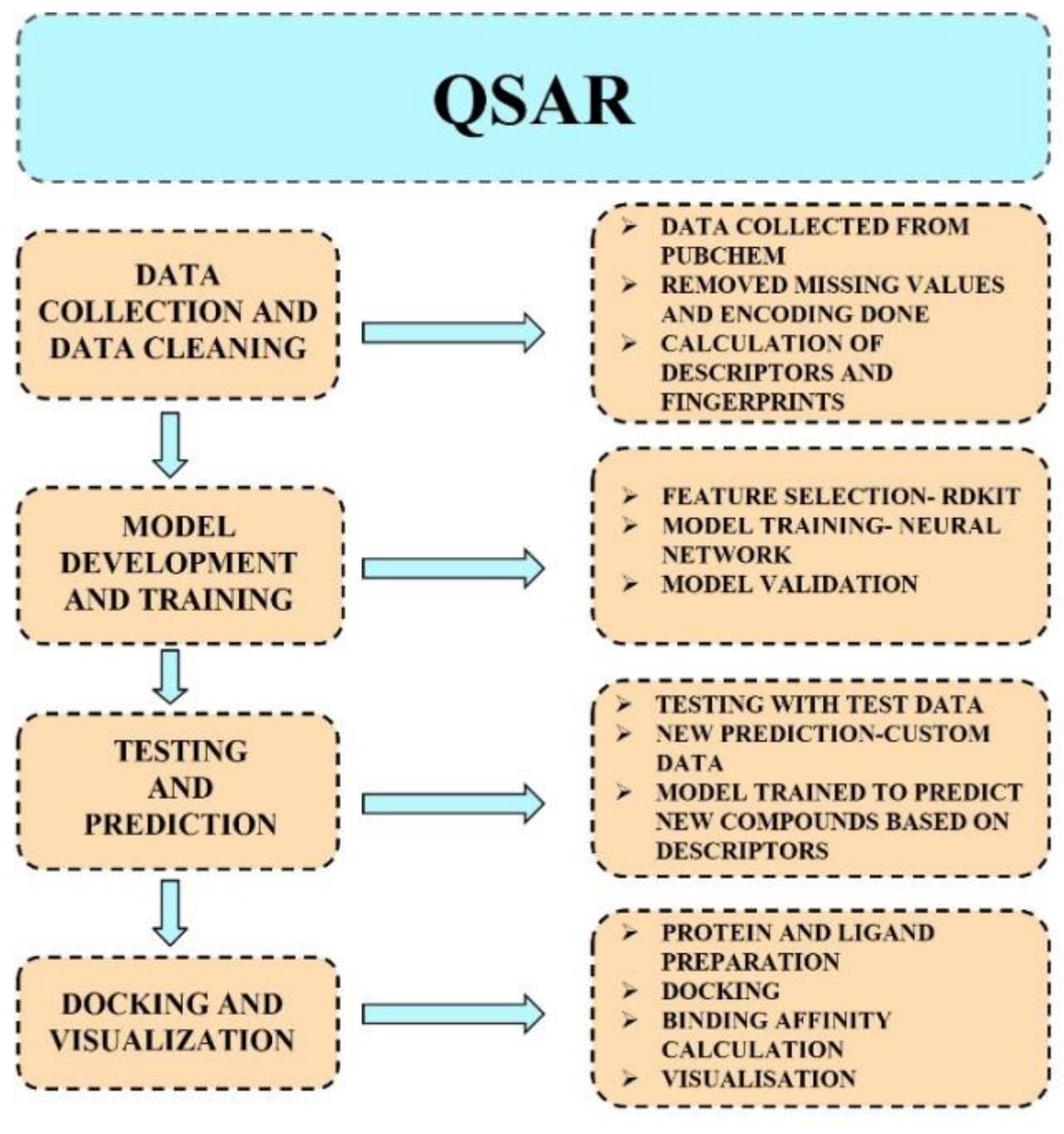

Fig. 1. Schematic representation of the QSAR modeling, molecular docking, and visualization methodology involved in the study. The figure was made using Microsoft® PowerPoint® 2021 MSO (Version 2508 (Build 19127.20264).

## III. Results and Discussion

### A. Dataset Characteristics

The initial modeling matrix comprised 232 chemical compounds. Examination of the raw activity data yielded 131 active compounds, 101 potentially inactive compounds, and 8 records classified as Inconclusive. The presence of this "Inconclusive" subclass introduces noise, which severely compromises the ability of a classification model to learn accurate structure-activity relationships for prioritization. Consequently, to ensure the fidelity and robustness of the model, the 8 inconclusive compounds were systematically excluded from the analysis. The rigorous cleaning process resulted in a final curated dataset, saved as “ER_DATASET_CLEAN.csv”, consisting of 224 compounds (131 Active and 93 Inactive), which established a definitive and reliable binary endpoint for the classification task (Fig. 2.).

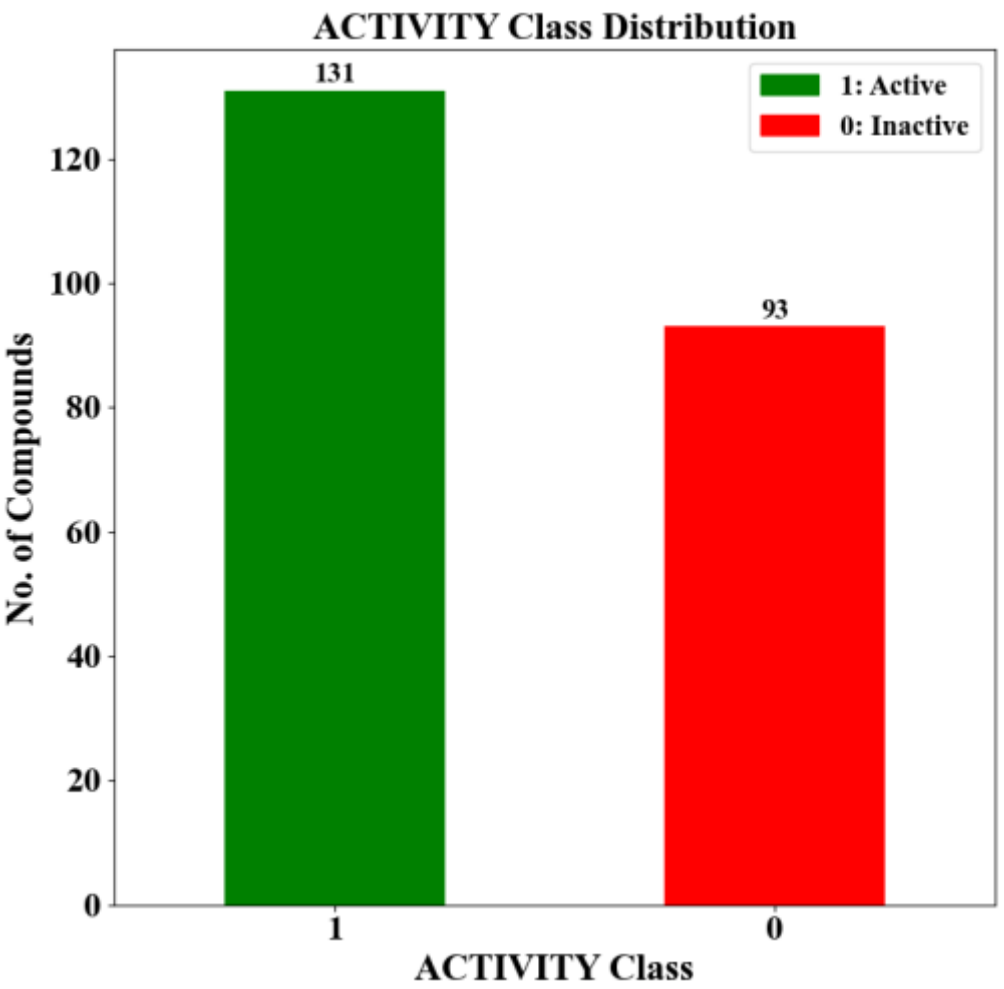

Fig. 2. Distribution of Activity Classes in the Curated Estrogen Receptor (ER) Binding Dataset. The bar graph illustrates the final class distribution within the curated dataset (n=224 compounds) used for training and validation of the deep learning QSAR model.

### B. Performance of the Model

The developed neural network demonstrated strong performance in classifying compounds as active or inactive. After 63 training epochs, the model converged with a final loss of 3.113, yielding accuracies of 96.65% on the training set and 91.30% on the test set, reflecting good generalization to novel data. The Receiver Operating Characteristic–Area Under the Curve (ROC-AUC) was 0.81, confirming its ability to reliably discriminate between the two categories. As summarized in Table I, the model achieved a precision score of 0.82 for both inactive and active compounds, minimizing false positives. Furthermore, it maintained a recall of 0.88 for the active class, highlighting its effectiveness in identifying highly active compounds.

TABLE I. Classification Report

| Class | Precision | Recall | F1-Score | Support |
|---|---|---|---|---|
| 0 (Inactive) | 0.82 | 0.74 | 0.78 | 19 |
| 1 (Active) | 0.82 | 0.88 | 0.85 | 26 |

### C. Docking of Predicted Molecules

From the screening of 100 novel compounds, the deep learning model identified 20 as potential actives. The reference ligand, Estradiol, showed a binding affinity of -10.79 kcal/mol, which was comparable to several predicted compounds when docked to the Estrogen Receptor (Table II). Binding affinity, expressed as the Gibbs free energy of binding ($\Delta G$), reflecting the strength of ligand–protein interactions, where more negative values correspond to stronger binding.

The docking results indicated that some predicted molecules displayed affinities on par with Estradiol, suggesting similar interaction potential within the receptor’s active site. For instance, CID:5870 achieved a binding affinity of –10.7 kcal/mol, while CID:5280961 scored –8.90 kcal/mol. The low standard deviation in results confirmed the robustness of the docking workflow. 3D visualization further showed that Estradiol and the predicted compounds occupied overlapping binding pockets (Fig. 3).

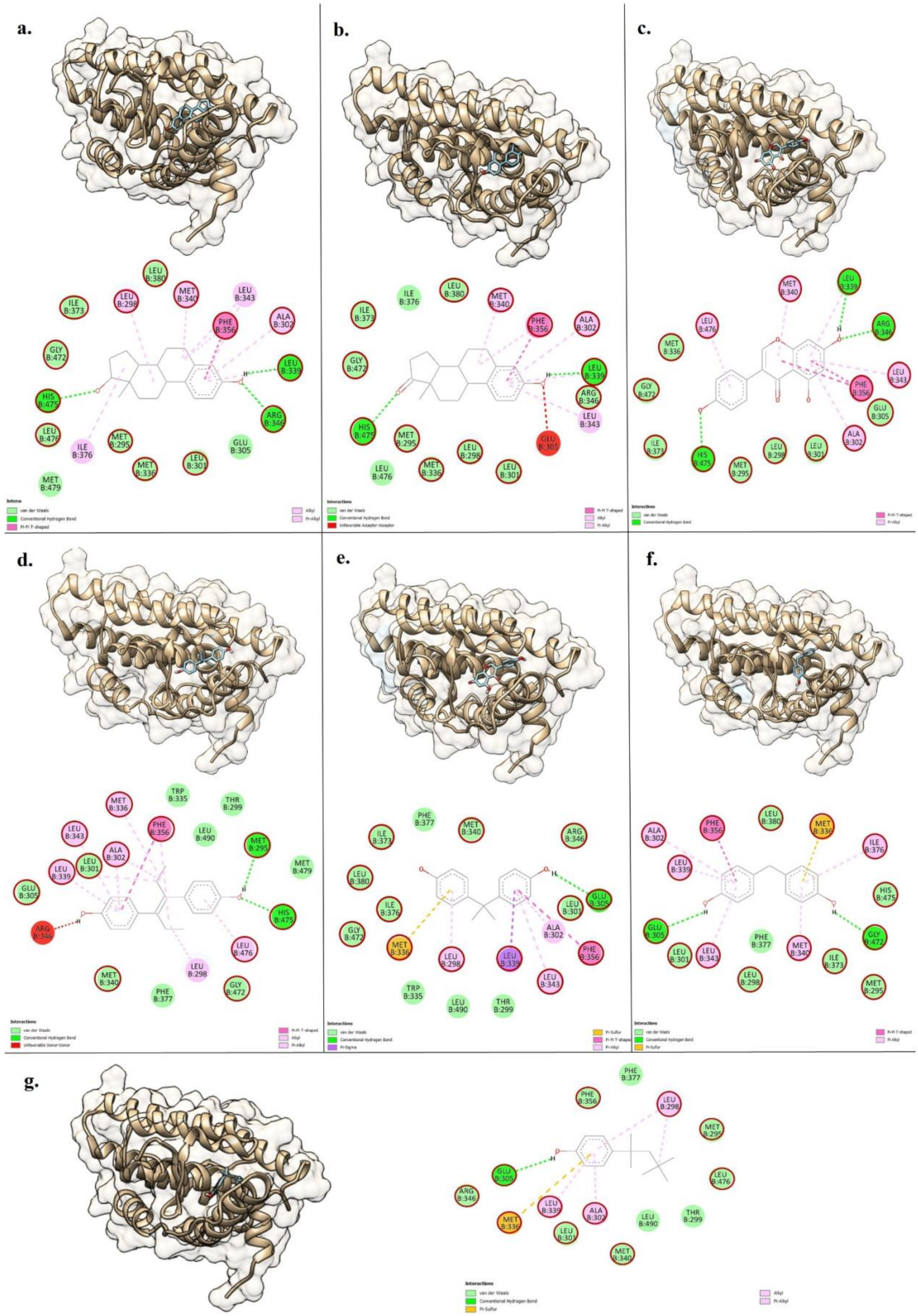

Fig. 3. 3D and 2D representations of docked complexes in the Estrogen Receptor protein. (a) Estradiol (CID:5757), (b) CID:5870, (c) CID:5280961, (d) CID:448537, (e) CID:6623, (f) CID:12111, (g) CID:8814

From the predicted dataset, the top 6 compounds identified were CID:5870 (environmental contaminant), Genistein [CID:5280961] (a phytoestrogen abundant in soy products and legumes), Diethylstilbestrol [CID:448537] (a synthetic estrogen once used therapeutically), Bisphenol A [CID:6623] (a common plastic monomer in polycarbonates), Bisphenol F [CID:12111] (an epoxy resin component and BPA alternative), and 4-tert-Octylphenol [CID:8814] (a degradation product of surfactants). These compounds are well-documented endocrine disruptors originating from diverse anthropogenic and natural sources [12].

2D interaction mapping revealed that both Estradiol and CID:5870 formed contacts with critical residues, including LEU380, MET340, PHE356, ALA302, LEU339, ARG346, LEU301, LEU298, MET336, MET295, HIS475, GLY472 and ILE373 (highlighted in Fig. 3). The overlap in residue (circled in red) interactions strongly suggests that CID:5870 may act through a mechanism analogous to Estradiol. Furthermore, CID:5280961 showed contact with amino acid residues, which includes, LEU476, MET340, LEU339, ARG346, LEU343, PHE356, GLU305, ALA302, LEU301, LEU298, MET295, HIS475, ILE373, GLY472, and MET336. Together, these findings indicate that the predicted inhibitors can effectively target the same binding site as the established ligand. This clearly shows the predictive efficacy of the developed model.

TABLE II. BINDING AFFINITIES AND PREDICTION PROBABILITIES OF TOP 6 PREDICTED COMPOUNDS

| Sl. No. | Prediction Probability | Compound ID | Binding affinity (kcal/mol) |
|---|---|---|---|
| 1 | 0.87 | 5870 | -10.7 |
| 2 | 0.75 | 5280961 | -8.9 |
| 3 | 0.87 | 448537 | -8.1 |
| 4 | 0.84 | 6623 | -7.8 |
| 5 | 0.84 | 12111 | -7.5 |
| 6 | 0.82 | 8814 | -7.0 |

## IV. CONCLUSION

This study demonstrates the utility of deep learning in predicting endocrine-disrupting chemicals with high accuracy and reliability. The developed DNN model, trained on a curated ER binding dataset, was able to generalize effectively to unseen compounds, as confirmed by molecular docking validation. Notably, several predicted compounds showed binding affinities and interaction patterns comparable to Estradiol, reinforcing the model's capacity to identify biologically relevant candidates.

By integrating computational toxicology with molecular docking, this work provides a cost-effective and scalable strategy for prioritizing environmental chemicals for further testing. Importantly, the implications extend beyond human health, EDCs disrupt reproduction and development across taxa, undermining population stability and ecosystem integrity. Thus, predictive approaches such as the one presented here can inform proactive regulatory decision-making and contribute to biodiversity conservation by reducing the risk of chemical-driven species decline.

Future research should expand these methods to additional hormone pathways and incorporate larger, more diverse datasets to further strengthen predictive performance. Ultimately, this computational framework represents a step toward more sustainable chemical management and the protection of both human and ecological health.